\documentclass[sigconf]{acmart}

\AtBeginDocument{%
  \providecommand\BibTeX{{%
    \normalfont B\kern-0.5em{\scshape i\kern-0.25em b}\kern-0.8em\TeX}}}

\setcopyright{none}
\copyrightyear{}
\acmDOI{}
\acmYear{}
\acmConference[W4A'24]{The 21st International Web for All Conference}{May 13--14,
  2024}{Singapore}
\acmBooktitle{Proceedings of the 21st International Web for All Conference,
 May 13--14, 2024, Singapore} 
\acmPrice{}
\acmISBN{}

\usepackage{balance}       
\usepackage{graphics}      
\usepackage[T1]{fontenc}   
\usepackage{mathptmx}
\usepackage{color}
\usepackage{booktabs}
\usepackage{textcomp}
\usepackage{makecell}
\usepackage{xspace}
\usepackage{fancyvrb}

\usepackage{amsthm}
\usepackage{hyperref}
\usepackage{subcaption}

\usepackage{microtype}        
\usepackage{ccicons}          

\newcommand{\eg}{\textit{e.g.}\@\xspace}
\newcommand{\ie}{\textit{i.e.}\@\xspace}
\newcommand{\etal}{\textit{et al. }}

\newcommand{\textnew}[1]{{\sffamily\bfseries#1}}

\renewenvironment{quote}
  {\list{}{\rightmargin=0.4cm \leftmargin=0.4cm}%
   \item\relax}
  {\endlist}

\usepackage{multirow}
\usepackage{enumitem}

\begin{document}
\title{``We are at the mercy of others’ opinion'': Supporting Blind People in Recreational Window Shopping with AI-infused Technology}

\author{Rie Kamikubo}
\affiliation{%
    \institution{University of Maryland, College Park}
}
\email{rkamikub@umd.edu}
\author{Hernisa Kacorri}
\affiliation{%
    \institution{University of Maryland, College Park}
}
\email{hernisa@umd.edu}
\author{Chieko Asakawa}
\affiliation{%
    \institution{Carnegie Mellon University}
    \institution{IBM Research}
}
\email{chiekoa@us.ibm.com}

\renewcommand{\shortauthors}{Kamikubo et al.}


\begin{abstract}
  Engaging in recreational activities in public spaces poses challenges for blind people, often involving dependency on sighted help.  Window shopping is a key recreational activity that remains inaccessible. In this paper, we investigate the information needs, challenges, and current approaches blind people have to recreational window shopping to inform the design of existing wayfinding and navigation technology for supporting blind shoppers in exploration and serendipitous discovery. We conduct a formative study with a total of 18 blind participants that include both focus groups (N=8) and interviews for requirements analysis (N=10). We find that there is a desire for \textit{push} notifications of promotional information and \textit{pull} notifications about shops of interest such as the targeted audience of a brand. Information about obstacles and points-of-interest required customization depending on one's mobility aid as well as presence of a crowd, children, and wheelchair users. We translate these findings into specific information modalities and rendering in the context of two existing AI-infused assistive applications: NavCog (a turn-by-turn navigation app) and Cabot (a navigation robot).
  
\end{abstract}



\maketitle

\section{Introduction}
Participation in recreational activities plays an important role in enhancing people's quality of life~\cite{iso1996leisure,pressman2009association,szczepanska2021covid,labbe2019participating}. Providing equal access to these pursuits is imperative for fostering social inclusion~\cite{hall2009social}, a key element in The United Nations Convention on the Rights of Persons with Disabilities affirming the rights to leisure and recreation~\cite{manca2017article}. However, blind people often experience limited recreational opportunities~\cite{salminen2014young,bandukda2020places,lieberman2023outdoor,rector2017enhancing}, especially those that take place in public spaces such as shopping~\cite{salminen2014young}. This is partially due to natural or built barriers and negative cultural attitudes~\cite{yu2015retail,stephens2020smooth,loi2017tourism,devine2015leveling,rimmer2006building}. 

To promote leisure, well-being, and social inclusion of blind people, in this paper we explore technological solutions for challenges typically faced by this community in recreational shopping. Specifically, the search aspect of browsing, also known as \textit{window shopping}, can be extremely difficult to perform without access to visual cues and awareness of the surroundings. We build on the rich literature that focuses on daily living activities such as wayfinding~\cite{sato2017navcog3,ren2023experiments,kacorri2016supporting,BlindSquare} and safety in navigation~\cite{katzschmann2018safe,li2016isana,kuribayashi2022corridor}
to elicit design recommendations that can inform these existing AI-infused\footnote{We adopt the term introduced by Amershi \etal~\cite{amershi2019guidelines} referring to ``features harnessing artificial intelligence (AI) capabilities that are directly exposed to the end user.''} systems to support recreational window shopping.

\textbf{Why window shopping?} Shopping is not only limited to acquiring consumer staples; it has been considered a meaningful activity that people engage in for a variety of personal and social reasons~\cite{ng2003satisfying,backstrom2006understanding,yu2015retail}. Many people choose to go window shopping as a form of recreation to learn about the latest trends, gain fashion or decorating inspiration, or socialize with friends~\cite{el2013shopping, bloch1989extending}. The serendipitous discovery of new information in window shopping also adds pleasurable experiences~\cite{bloch1991leisure,xia2010examination}. While our attitudes and experiences of shopping changed during the COVID-19 pandemic, we don't know how long this impact will last~\cite{diaz2023covid}. Sheth~\cite{sheth2020impact} argues that ``when an existing habit or a necessity is given up, it always comes back as a recreation or a hobby,'' and wonders whether shopping can ``become more an outdoor activity or hobby or recreation.'' Others also see hope for a post-pandemic retail renaissance~\cite{danziger2020malls, weintraub2020future} aligned with the recent trends~\cite{snider2023death, rothenberg2023us}. Thus, the need to make window shopping inclusive remains. 

To inform the design of existing wayfinding and navigation systems for enhancing recreational window shopping opportunities, we conduct a formative study with blind participants (N=18).  First, we gain insights from two in-person focus groups (4 participants each) on window-shopping challenges and opportunities, and identify a set of information needs of blind shoppers. Then, we conduct remote one-on-one interview sessions with a second group (N=10) to assess the types of notifications blind people prefer. In these sessions, participants are grounded in a simulated scenario of future guidance systems (\eg, walking with guide robots) that support both obstacle avoidance and information access for recreational shopping. 

Our findings reveal a desire for push notifications of promotional information (\eg, window displays), whereas pull notifications of detailed shop information (\eg, general fashion styles or items carried, target age groups) only for shops of interest. Participant feedback indicates that information about obstacles and points-of-interest (POIs) needs customization depending on their mobility aid, presence of a crowd, children, and wheelchair users. We discuss how these findings can be translated into specific information rendering and modalities in the context of two AI-infused assistive applications: a guide robot (\ie, Cabot~\cite{guerreiro2019cabot}) and a turn-by-turn navigation mobile app (\ie, NavCog~\cite{sato2017navcog3}).

The main contribution of this work is empirical. By gathering and analyzing information requirements with blind people through a combination of focus groups and one-on-one interviews, we present design implications for future guidance systems to take into account the requirements shaped by the specific user needs and preferences in the context of recreational shopping. We also see how our study provides insights that are applicable to a broader context, such as ``The Future of Shopping Centers'' released in 2018 by Kearney~\cite{brown2018future}. This remains relevant to many later articles suggesting technological innovations for post-COVID in-person shopping and retail experiences~\cite{moore2023future,weintraub2020future,reese2020integrating,brown2021killing}, with the potential for AI in this space~\cite{guha2021artificial}.

\section{Related Work}
This research aims to improve access to recreational opportunities for blind people. We draw on two areas of related work in accessibility: research with the blind community on recreational activities in public spaces, with a particular focus on shopping scenarios, and research on wayfinding and navigation technology for blind people that can be applied in these scenarios. 

\subsection{Accessible Recreational Activities}
For many blind people, mobility and information access challenges can limit their recreational pursuits in public spaces, whether indoor or outdoor~\cite{salminen2014young,lieberman2023outdoor,lieberman2002self,rector2017enhancing}. There are additional barriers to participation in terms of transportation and high dependence on others' assistance~\cite{lieberman2002self,rector2015exploring}. In response, we see efforts to understand and remove these barriers in recreational activities, including travel~\cite{stephens2020smooth,devile2020accessible,muller2022traveling}, nature~\cite{burns2009inclusive, burns2008exploring,bandukda2020places}, and sports~\cite{rimmer2005conspicuous,rector2015exploring,jaarsma2014barriers}. Nevertheless, acquiring visual information is often a prerequisite for recreational experiences, such as in museums and art galleries. To foster inclusive museum experiences, various researchers have explored the accessibility of exhibitions and artworks, via guided tours, audio descriptions, or tactile cues~\cite{kwon2019supporting,ahmetovi2021touch,reichinger2016gesture,holloway2019making,iranzo2019exploring}. Technology to support navigation in museums for blind people has also been an important area of exploration~\cite{jain2014pilot,wang2022bento,kayukawa2023enhancing}.

Shopping, the focus of our work, is another activity that offers considerable recreational potential~\cite{backstrom2006understanding}, particularly window shopping~\cite{bloch1989extending,xia2010examination}. Shopping malls facilitate this experience as entertainment venues~\cite{el2013shopping,farrag2010mall}, and their accessibility could hold various implications for blind people despite the growth of online shopping~\cite{baker2006consumer,baker2002can}. Baker \etal~\cite{baker2006consumer} suggest that beyond meeting recreational needs, blind and low vision consumers derive a sense of belonging and presence within their community through in-person shopping experiences. To explore ways to make these experiences accessible, previous research has investigated the challenges faced by this consumer group concerning retail layout, purchasing methods, and environment barriers within malls~\cite{kameswaran2019cash,bradley2000study,baker2002can,swaine2014exploring,yu2015retail,tullio2021you}. 

We see extensive research on technological solutions for shopping for blind people from a literature review~\cite{elgendy2019making}. However, it is often discussed in the context of daily activities, such as grocery shopping~\cite{yuan2019constructing}. Many early efforts focus on supporting navigation inside the supermarket and identification of desired items on shelves, often using the infrastructure of RFID systems with barcode scanning~\cite{alnfiai2014virtualeyez,lanigan2006trinetra,lopez2011blindshopping,nicholson2009shoptalk}. With advances in computer vision and mobile technologies, researchers have introduced new interaction methods for shopping~\cite{elgendy2019making}, including wearable devices offering audio guidance to supplement visual information of consumer staples~\cite{zhao2016cuesee,orcam,boldu2020aisee,wang2023mobieye,zientara2017third} and applications that indicate clothing patterns and colors~\cite{yang2014assistive,stearns2018handsight,stangl2018browsewithme,stearns2018applying}. 

Despite the large body of work on accessible shopping for blind people, supporting recreational window shopping remains unexplored, with a few exceptions to test wayfinding and navigation systems in shopping malls (\eg,~\cite{sato2017navcog3,saha2019closing}). Yet, these efforts are intended to support blind people to reach their destination effectively. While this is still an important research direction, Guerreiro \etal~\cite{guerreiro2019airport} investigating airport navigation for blind people found a significant limitation in their travel experiences even when supposedly \textit{assisted} -- they are escorted to the target gate but have no opportunity to explore nearby shops or restaurants, reflecting their desire to ``get up and move around.'' Exploration and serendipitous discovery of new information, as in window shopping, is deemed a challenge for blind people.

\subsection{Accessible Wayfinding and Navigation}

There has been a wide array of research on orientation \& mobility (O\&M) for blind people beyond directional guidance~\cite{thoo2023large}, such as informing nearby points of interest (POIs)~\cite{kacorri2016supporting,Soundscape,gleason2018footnotes} and obstacles~\cite{presti2019watchout,liu2015isee,katzschmann2018safe}, with various form factors being investigated~\cite{real2019navigation,tapu2020wearable,khan2022recent}. We expand on specific form factors used in these wayfinding and navigation systems that have the potential to be applied in recreational window shopping scenarios: smartphones and wearable devices and robots. 

\textit{Smartphones and Wearable Devices.} Smartphone-based solutions have been common in previous research, allowing researchers to leverage various sensors and technologies for user localization and implement turn-by-turn instructions to reach a target destination~\cite{sato2017navcog3,murata2018smartphone,kuribayashi2022corridor,cheraghi2017guidebeacon,kim2016navigating}. These solutions often come with conveying the location of nearby POIs, helping users build a rich awareness of their surroundings~\cite{BlindSquare,kacorri2016supporting,Soundscape}. In particular, Microsoft's Soundscape smartphone app~\cite{BlindSquare} provides 3D audio cues tied to features in the real-world environment to provide information about POIs in the user’s vicinity. Gleason \etal extend this design by augmenting spatial audio with rich textual descriptions of POIs for additional functional or visual context.

Another line of research focuses on detecting obstacles and providing auditory or tactile feedback to help blind users avoid them, often using wearable devices~\cite{dakopoulos2009wearable,katzschmann2018safe,li2016isana,rodriguez2012assisting}. These systems have the potential for application in minimizing barriers in shopping environments \eg, obstacles in the open aisles of the malls~\cite{swaine2014exploring}. Yet, designing appropriate feedback for obstacles often remains complex when it can increase confusion and cognitive load~\cite{sato2019navcog3}.

\textit{Robots.} The development of guide robots for blind people has been an active area of research for decades~\cite{tachi1984guide, galatas2011eyedog,nanavati2018coupled,azenkot2016enabling,guerreiro2019cabot}. One of the key benefits of robot guidance is obstacle avoidance. By physically grasping a robot that guides and influences their trajectory, blind users can gain enhanced mobility and safety~\cite{tobita2017examination,nanavati2018coupled,tachi1984guide,megalingam2019autonomous,galatas2011eyedog}. In such interaction, the feedback regarding obstacles can be minimized to prevent overwhelming the user. Instead, guide robots can broaden their functionality to provide additional information related to POIs and landmarks~\cite{guerreiro2019cabot,tobita2017examination,lacey2000context}. 

We see that building on the work of wayfinding and navigation systems for blind people opens potential avenues for recreational window shopping opportunities, especially with guide robots that can support both obstacle avoidance and provision of environmental information~\cite{guerreiro2019cabot}. However, there is still a limited understanding of design features to apply these systems in recreational shopping scenarios. To our knowledge, there is no comprehensive study that investigates the requirements in this context. For example, what are the relevant POIs in recreational window shopping? What kind of information do blind users want to access and control for enhanced recreational shopping experiences? We aim to address these questions in our study.
\section{Study Design}

With our goal to inform the design of AI-infused systems to ultimately support recreational window shopping, we conduct a formative study with blind participants (N=18). The study is broken into two parts: (i) in-person focus groups involving 8 blind participants and (ii) remote one-on-one interview sessions with a new group of 10 participants. Section~\ref{sec:procedure} outlines the study details. We obtained (IRB-approved) informed consent and compensated participants for their time and expertise at a rate of \$25 per hour.

\subsection{Participants}
Participant recruitment for this study took place in the United States. We recruited participants through an existing mailing list in our lab, consisting of blind people from the local community who had previously participated in studies conducted by our research group, and word of mouth. Table~\ref{tab:participant} shows the demographic information of all blind participants, with a noticeable gender skew observed towards women. While this may be partly due to the nature of our convenience sampling approach, we can presume motivation factors by gender influencing participants' involvement in our study~\cite{lakomy2020motivation}. Yet, in hindsight, our study surfaces perspectives from groups that are often underrepresented within HCI~\cite{offenwanger2021diagnosing} and accessibility~\cite{kamikubo2022data}. 

In the 90-minute focus group sessions, we invited 4 participants per session (Group 1: P1 - P4, Group 2: P5 - P8). Participants were 39 to 78 years old (median=66, IQR=70.3-60.3). Half had a guide dog. The other half used only the white cane. While we sampled the participants for each group based on their schedule availability, all Group 1 members mainly do in-person shopping, whereas all Group 2 members described their preferred choice between online and in-person shopping depending on the type of merchandise (\eg, in-person for clothes/shoes, online for gifts/electronics). 

In the 45-minute one-on-one sessions, we invited 10 participants (P9 - P18) with ages ranging from 29 to 72 years old (median=45, IQR=51-36.8). For half of the participants, the white cane was the primary mobility aid. One of the participants previously used a white cane but since transitioning to a wheelchair, they relied on sighted help. The other 4 participants had a guide dog. Six participants (P9, P10, P11, P13, P16, P18) reported going for in-person shopping (\eg, in shopping malls) about 4 to 6 times a year. The other participants (P12, P14, P15, P17) indicated 1 to 3 times a year. These frequencies are much lower considering the average mall visits in the U.S. (\ie, 3.1 times per month in 2018)~\cite{lindner2023gitnux}. Yet, all participants, except P11, indicated that they like to do in-person shopping as a recreational activity, especially for clothes or shoes. P11 expressed a similar preference for in-person and online shopping, as the online shopping experience provides a \textit{``sense of better independence.''} 

\begin{table}[t]
\small
\caption{Participant demographics from focus groups and interviews including gender, age, mobility aid, and vision level.}
\centering
\begin{tabular}{ | >{\centering\arraybackslash}m{0.19\linewidth} | >{\centering\arraybackslash}m{0.05\linewidth} | >{\centering\arraybackslash}m{0.1\linewidth} | >{\centering\arraybackslash}m{0.05\linewidth} | >{\centering\arraybackslash}m{0.18\linewidth} | >{\centering\arraybackslash}m{0.17\linewidth}|}
\hline
Study Session & ID & Gender & Age & Mobility Aid & Vision Level \\
\hline
\multirow{4}{*}{\vtop{\hbox{\strut Focus Group}\hbox to 1.5cm{\hfil 1 \hfil}}}&P1 & woman & 67 & guide dog & legally blind\\
    \cline{2-6}
    &P2 & woman&69 & white cane & totally blind\\
    \cline{2-6}
    &P3 & woman&74 & white cane & totally blind\\
    \cline{2-6}
    &P4 & woman&78 & white cane & totally blind\\
    \hline
\multirow{4}{*}{\vtop{\hbox{\strut Focus Group}\hbox to 1.5cm{\hfil 2 \hfil}}}&P5 & woman&39 & guide dog & totally blind\\
    \cline{2-6}
    &P6 & woman&46 & guide dog & totally blind\\
    \cline{2-6}
    &P7 & woman&65 & guide dog & totally blind\\
    \cline{2-6}
    &P8 & woman&65 & white cane & legally blind\\
\hline
\multirow{10}{*}{\vtop{\hbox{\strut One-on-One}\hbox {\strut Interviews}}}&P9 & man&29 & white cane & totally blind\\
	\cline{2-6}
    &P10 & woman&35 & guide dog & totally blind\\
    \cline{2-6}
    &P11 & woman&36 & guide dog & totally blind\\
    \cline{2-6}
    &P12 & man&39 & wheelchair \& sighted help & legally blind\\
    \cline{2-6}
    &P13 & man&44 & guide dog & totally blind\\
    \cline{2-6}
    &P14 & woman&46 & guide dog & totally blind\\
    \cline{2-6}
    &P15 & woman&48 & white cane & totally blind\\
    \cline{2-6}
    &P16 & woman&52 & white cane & totally blind\\
    \cline{2-6}
    &P17 & woman&65 & white cane & totally blind\\
    \cline{2-6}
    &P18 & woman&72 & white cane & totally blind\\

\hline
\end{tabular}
\label{tab:participant}
\end{table}

\subsection{Procedure}
\label{sec:procedure}
Focus group participants were engaged in a free discussion centered around strategies and coping mechanisms to perform recreational shopping, the main challenges of navigating public places such as shopping in malls, and information needs to enjoy window shopping. 
Specifically, the session started with the research team clarifying the study's objectives. The team then provided examples of state-of-the-art navigation technologies with a focus on those that have a physical form factor, such as guide robots that could be extended to support window shopping. To facilitate discussion on the information needs and technological opportunities for blind window shopping, the groups were also instructed on what some of these systems (\eg, NavCog~\cite{sato2017navcog3}, Cabot~\cite{guerreiro2019cabot}) can offer in terms of alerting or avoiding obstacles and informing POIs to the user. Via a turn-taking method, participants were then invited to share their shopping experiences that did not include grocery shopping, whether done in person or online, alone or with someone. To provide more context, participants included any rationale and preferences around their shopping activities. Last, the groups were given a specific scenario: \textit{to imagine walking with a guide robot like the one discussed in the study inside a shopping mall}. Continuing with a turn-taking method, participants discuss their information needs grounded in this concrete scenario. 

With our initial information requirements synthesized from the focus groups, we proceeded to conduct in-depth interviews with a third group comprising 10 blind participants. This session aimed to analyze and refine these requirements. Grounded in the same scenario of walking with a guide robot inside a shopping mall, participants were prompted with a list of potential features extracted from the focus groups, divided into three sections: (a) obstacles, (b) POIs, and (c) shops. For each section, participants indicated whether and how they would prefer accessing features of interest in terms of pull and push notifications. For push notifications, participants could indicate whether they wanted toggle controls \eg, turn them off in a given context. For pull notification, participants could indicate the modality of interaction \eg, asking verbally or pressing a button. Participants also identified any missing features in the list to expand on their information needs. 

The results of the one-on-one interviews are the counts of informative notifications related to shops, POIs, and obstacles, as well as qualitative data about their attitudes to accessing particular features of the surroundings to enhance window shopping opportunities in the future. We thematically analyzed the qualitative study data obtained from all sessions that were recorded and transcribed.

\subsection{Reflections of Our Study Design}
\label{reflections}
Our original study protocol following the focus groups was meant to have participants test a simulated scenario of being led by a guide robot inside a shopping mall. An experimenter would simulate the response of the robot using a mobile app that conveys relevant audio messages about the surroundings to the participant walking next to the experimenter. The app was built for an in-person experiment  and had UI elements such as buttons and dropdowns to read pre-defined information stored in a dictionary according to participants' commands through voice or movements. For instance, the participant could choose to give a verbal `Read' command to pull promotional information (\eg, `Spring Sale') or automatically receive push notifications as they slow down or stops near the entrance of the store of interest. There was also a `Search' command to hear information about certain areas/services in the mall. The app had a function to store logs of tapped UI elements for desirable information and included an input field for any additional information that did not exist in the dictionary.

Unfortunately, the COVID-19 pandemic during our data collection forced us to cancel the in-person portion of this study. Instead, participants were invited for interviews through one-on-one Zoom calls. The app was still used but merely served as a demo of what one would hear in an imagined window shopping scenario in an indoor mall with a floor of retailers located on both sides. We provided a detailed description of the shopping mall and shopping activities, such as passing by retail stores, window displays, signs, potential obstacles, and other pedestrians, to better contextualize the assessment of information needs.

\section{Focus Groups}
The results from the focus groups provide data on blind participants' perspectives, current strategies, and challenges in recreational shopping. To enhance their future window shopping opportunities, we also surface various information needs to formulate initial information requirements. 

\subsection{Engaging in Recreational Shopping}

\textbf{In person vs. online.} When discussing their in-person and online shopping experiences, all participants from Group 1 preferred in-person recreational shopping as it creates rich sensory experiences that are missing from online shopping: 


\begin{quote}
\textit{``When you're doing in-person shopping and when you have the items in your hand, you're using all the senses that you can use in replace for vision, whereas if you're dealing with online shopping, you are not using all of the senses. I think it's really good to be able to touch and feel.''} - (P4)
\end{quote}

Group 1 also highlighted accessibility concerns with online shopping platforms, strengthening their preference for in-person recreational shopping. They often have to rely on alt-text descriptions that lack information for a better conceptualization of product selections:

\begin{quote}
    \textit{``I try to do some sort of window shopping online. It is so difficult even to find the link that is going to tell you the descriptions...Every site is different and then sometimes the descriptions are so lacking in information even once you read it, you still don't know what that is going to look like.''} - (P1)
\end{quote}


Group 2, on the other hand, had a detailed discussion on how online and in-person shopping are preferred depending on the type of merchandise to browse. Online shopping was common for gifts or electronics where they could rely on reviews, whereas in-person shopping was preferred for clothes or shoes, especially for special occasion items (\eg, event gowns): 



\begin{quote}
\textit{``I have a niece and nephew and I bought them gifts a lot. And usually they are online because I can't read those reviews I like reading reviews you know I read like twenty or thirty reviews before I bought them, just to make it's exactly what I think it is. So shopping online is easier like that. I agree about shoes [to shop in person]. If I'm also buying something fancy if I have to buy something that I have to wear to events, then in-person is ideal. But then again I go with people.''} - (P8)
\end{quote}


Two participants (P5, P6) also commented on their preference for in-person recreational shopping as a form of \textit{``social trips''} (P5), emphasizing that they would not engage in it alone:


\begin{quote}
    \textit{``My husband and I just took a trip...We do like `window shopping' when it's nothing that we have to buy. We went to a bunch of wood shops and antique shops and we bought nothing but that was fun. Because he's fully sighted, it's like 'let's look at this, check that out.' We did it together but I never would have done that on my own. It becomes not fun.''} - (P6)
\end{quote}

\textbf{Assistance from others.} In addition to discussing their preferences, both groups delved into their strategies for in-person recreational shopping. All participants mentioned that they typically have sighted companions like friends or family members. They expressed their need for sighted help for various shopping tasks, from obtaining information about product selections to finding their way to specific locations within stores:

\begin{quote}
    \textit{``I always go with somebody if I'm going to buy clothing because I want to know if something looks good, in addition to of course finding out whether it fits and so that means I need to get to a fitting room and all of these kinds of things that are not very easy to do without another person.''} - (P3)
\end{quote}

Despite seeking sighted companions for recreational shopping, some participants (P1, P2, P6, P8) found it challenging. They instead ask for assistance from sales clerks for product descriptions or mall security guards to navigate between stores. P1 and P2 specifically reported their successful experience with a free personal stylist service at a department store that helped them explore selections of clothes. These choices about whom to approach for assistance could be explained by prior literature~\cite{thieme2018everything}, which suggests that blind people consider the social costs of requesting favors from close connections (\ie, friends) and often prefer to receive assistance from professionals tasked with assisting. 

Two participants (P2, P7) reported another strategy that could mitigate these social costs of ``appearing to be dependent''~\cite{brady2013investigating}. They search online for information about malls or stores before visiting in person with others, showcasing the potential of virtual navigation to support this approach (\eg,~\cite{guerreiro2017virtual}). Yet, they still expressed the need for sighted help to receive additional information beyond what they found in their prior online searches: 

\begin{quote}
    \textit{``I look on the internet to try to find out about the stores [in the mall] but you can't always find out what is everywhere. Sometimes if I'm with other ladies, they will tell me what they have...So you have to ask people and they will tell you because there are stores you never heard of.''} - (P2)
\end{quote}

\subsection{Lack of Opportunities for Window Shopping}

Both groups discussed their limited opportunities for window shopping \ie, browsing stores in shopping malls. One major contributing factor was the lack of independent access to information. Many participants (P1, P3, P4, P6, P7) shared their frustration while seeking and receiving information from others:

\begin{quote}
    \textit{``The part of the frustration is that of a line of questioning. My husband will say something and then I'll ask a quick question. He is not thinking as fast as I am about the items that I want to look at. Friends are a bit different, knowing what I want to look at. We are at the mercy of others' opinion.''} - (P7)
\end{quote}

All participants thus agreed that having specific target stores or items in mind is essential to receiving the kind of information they
desire from people. Otherwise, they constantly need to depend on others and be informed about their surroundings:


\begin{quote}
    \textit{``When I go to the mall, I want to know what is there. When I go with a friend, can you tell me what stores we are passing? Any blind person really needs to know where she wants to go.''} - (P4)
\end{quote}



In addition, two participants (P5, P6) highlighted issues of ``sighted people interference'' preventing them from engaging in window shopping. They expressed worries about how their actions might be interpreted by retail staff or other shoppers, similar to the concerns as blind people negotiate their actions in the social context (\eg, attending to other pedestrians' movements to avoid being seen as `rude' if accidentally walking close to them~\cite{thieme2018everything}):

\begin{quote}
    \textit{``One of the things that I think inhibits me from sometimes shopping in person is in terms of people. If I could just walk around the store and touch stuff and not worry about someone coming over and being offended or coming over and like `what are you doing there?' or me running into somebody, I would be a lot more inclined to do that. Because of sort of the sighted people interference, it's so hard.''} - (P6)
\end{quote}


\subsection{Information Needs in Window Shopping}
\label{features}

Participants discussed the important features of the surroundings that could enhance their window shopping opportunities in the context of walking inside a shopping mall with a guide robot designed to notify these features to the user. Accompanied by a demo of a related system (\ie, Cabot~\cite{guerreiro2019cabot}), they were prompted with three key topics to guide their discussions: (a) obstacles, (b) POIs, and (c) shops. We detail their information needs along with the feedback they provided.

\textbf{Notifying obstacles.}
Guide robots can notify users about obstacles on their path during navigation (often via audio messages \eg, \textit{``avoiding a person''}), but the relevance of such feedback tends to be context dependent~\cite{guerreiro2019cabot}. In our focus groups, we quickly recognized different preferences for notifications about obstacles. Participants often found information related to \textnew{pedestrians} as unnecessary, with a few exceptions (P3, P4) of wanting to be notified of crowded areas---situations that could pose a particular threat~\cite{branham2017someone}. Many still preferred to minimize these notifications, partly because they could rely on their ability to detect other pedestrians or the crowd, as \textit{``it will usually be noisy''} (P2). Similarly, while participants discussed \textnew{floor changes} such as wet floors or carpets as potential obstacles, only one participant (P2) cared for such information. Most participants preferred to use their mobility aids and/or senses to detect them. These preferences are perhaps reflective of blind users' concerns of losing spatial awareness when the amount of guidance is overwhelming~\cite{sato2017navcog3}. In fact, a few participants (P1, P5) alerted how such notifications could result in information overload:

\begin{quote}
\textit{``No no because I could hear that [pedestrians]. That would be constant in a shopping mall...Get too close to the food court, it would get much louder.''} - (P5)
\end{quote}

In continuation of their discussion on whether a guide robot should notify obstacles, participants identified certain objects of interest, especially those that are \textit{``functional''} (P2). For example, both groups referred to \textnew{trash cans}, \textnew{seating areas}, \textnew{displays}, and \textnew{elevators} as potential obstacles found in the middle of the shopping mall floor. They were on common ground that these obstacles should be notified depending on individual needs, as \textit{``different people need different things''} (P3). One idea was to pre-set objects of interest in settings and easily toggle notifications on or off for these objects:

\begin{quote}
    \textit{``There are times that you may want to find a chair to sit down in the mall but then there are times when you want to get from A to B place, and you don't really care about all the obstacles. So the best thing for me would be to have an option to turn on or off.''} - (P1)
\end{quote}


The need for configuring notifications also became apparent for \textit{non-functional} obstacles in the environment, such as \textnew{columns} or \textnew{planters}. Often, this was influenced by participants' mobility aids and navigation practices. Among participants using the white cane, one common preference was to be notified of these obstacles, serving as helpful cues for orientation~\cite{williams2014just}:

\begin{quote}
    \textit{``I think there's a fine line between not enough information and maybe too much information but I would like that, you know. One of the problems I always had with my note taker for points of interest was that you could get from place to place but you don't know what you're passing along the way...The same thing applies to the mall. You don't know what you don't know.} - (P8)
\end{quote}

In contrast, those navigating with a guide dog typically don't question the obstacles around, as guide dogs are taught to find a safe path forward~\cite{lloyd2008guide}. All but P5 guide dog owners in our focus groups referred to the same expectation when walking with a guide robot: 

\begin{quote}
\textit{``If it's gonna walk me around the planter then because I'm so used to the dog walking me around things and if the guide dog makes a sudden move, you just go. That's what guide dogs do.''} - (P6)
\end{quote}




\textbf{Notifying POIs.} POIs are places that might interest users during navigation~\cite{sato2019navcog3}. After discussing in detail about obstacles, participants expanded on POIs in the shopping mall environment aside from shops. These included \textnew{restrooms}, \textnew{water fountains}, \textnew{cafe}, \textnew{information counters}, and \textnew{ATMs}, which all aligned with semantic features found in navigation apps for blind people (\eg, NavCog~\cite{sato2017navcog3}). They found such \textit{incidental} information useful, as it is typically missing from their interactions with their environments:

\begin{quote}
    \textit{``It's good to know things like... `There's a restroom on your left.' `There's this cafe on your right.' Those things are really helpful. When you're walking past cafe or whatever, that's great. It's just sort of incidental information that you would get if you are a sighted person.''} - (P6)
\end{quote}

\textbf{Notifying shop details.} Last, both groups discussed their information needs about shops to enhance their window shopping opportunities. All participants agreed that it is important to automatically hear \textnew{shop names} as they walk past the shops and have control for \textnew{more detail}, such as through \textit{``a button on the robot''} (P6, P8) to access a second layer of information. For more details, they articulated notifications about \textnew{shop categories} such as men's, women's, or teen's fashion to know the general items or styles the shops carry. Group 2 also discussed ways to notify price ranges, including categorization as \textit{``fancy''} (P6) or parameters by \textit{``dollar signs''} (P7). Participants (P2, P3, P4, P8) showed additional interest in promotional information such as \textnew{sales}, \textnew{new arrivals}, or \textnew{grand opening}. With various layers of information about shops, establishing a clear information hierarchy was deemed important:

\begin{quote}
    \textit{``I want to have a way to finding out more information if I need it...I need to know the names of all the stores that I'm passing and then when I get to any given store I would like to know, number one what that store has to offer me and number two if it has any special signs such as sales or clearance.''} - (P3)
\end{quote}



Some participants (P1, P2, P8) also showed interest in knowing \textnew{window displays}, such as the styles of clothes on mannequins. Such visual information was found useful to better understand the types of merchandise offered by the shops:

\begin{quote}
    \textit{``Often if I'm going shopping and say it's the sort of the beginning of the season, I will say to the clerk, `what's the color of this year,' `what are the main colors of this year' just to get an idea of what I might expect a lot of.''} - (P1)
\end{quote}


Yet, there were still comments by others (P3, P4, P5) questioning the usefulness of knowing what is on the window display since it is mostly visual information that \textit{``we can't really experience''} (P5). One possible explanation for their perception of window displays could be linked to the onset age of blindness. Experiences for aesthetic enjoyment of congenitally blind people can differ from those who have become blind later on in life~\cite{li2023understanding}. Indeed, one of the participants (blind since birth) showed no interest in information intended for aesthetic purposes:

\begin{quote}
    \textit{``If the stuff in the display is something that is in the shop that I am going to buy, then it is interesting. If the stuff in the display is something to make it beautiful, like artificial flowers, then I don't need to know.''} - (P3)
\end{quote}

\section{Requirements Analysis}
Through one-on-one interviews, we assessed the initial information needs synthesized from the focus groups. To inform the future design of user interfaces to support blind people's window shopping experiences, we solicited participants' preferences for accessing information, guided by the following classification of notification attributes derived from the focus group insights: 

\begin{description}[leftmargin=*]
    \item[Push notify.] User is automatically notified of features of the surroundings \eg, restrooms and shop names.
    \item[Pull notify.] User requests information relevant to features of the surroundings \eg, promotional information for the nearby shops. 
    \item[Pre-set notify.] User pre-sets specific features of interest (\eg, chairs \& trash cans) to toggle notifications on/off based on individual needs.
    \item[No feedback.] User is not interested in information such as pedestrians; this is especially the case for most guide dog owners.
\end{description}

Below, we report our analysis of the assessment which revealed different types of notifications related to obstacles and POIs (in Section~\ref{notifyom}) and shops (in Section~\ref{notifyshops}), specifying the rendering and modality to access relevant information. 

\subsection{Types of O\&M Notifications}

\label{notifyom}

\begin{figure}[!th]

\begin{subfigure}{1\linewidth}
  \centering
  \includegraphics[width=1\linewidth]{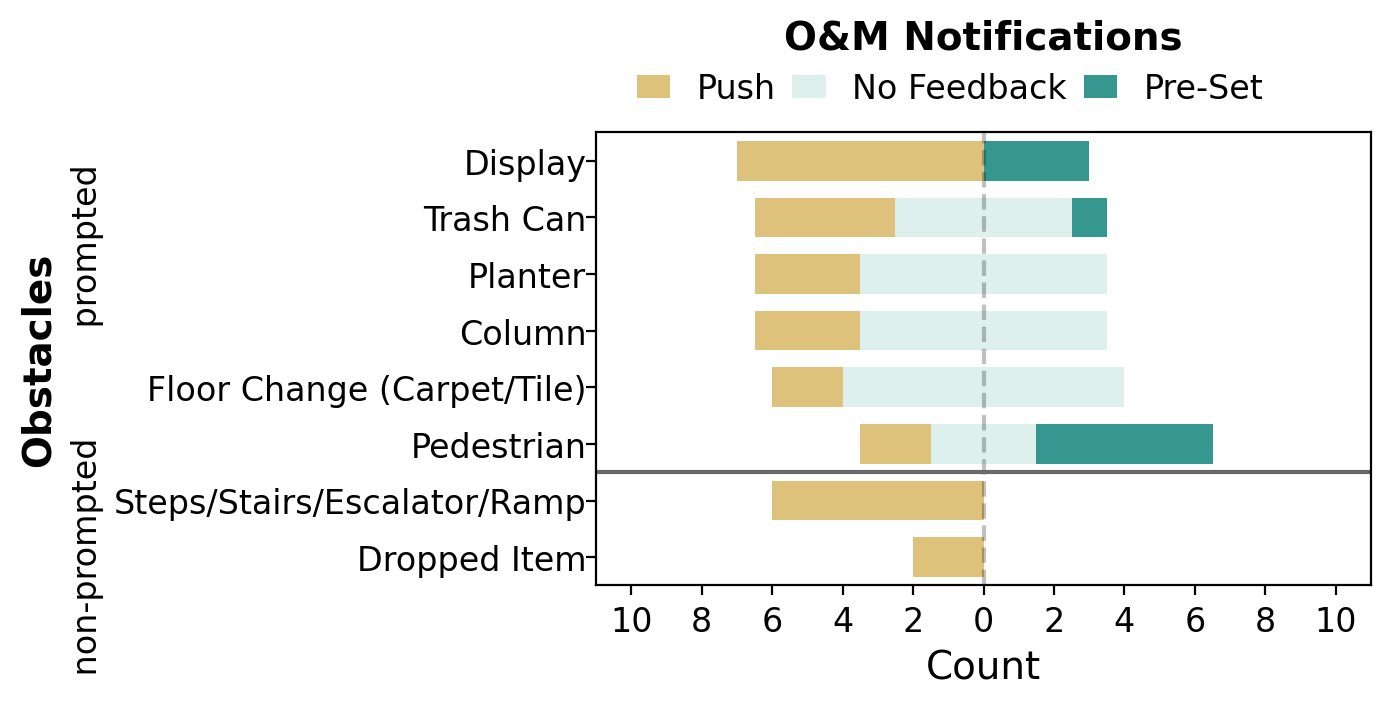} 
  \caption{}
  \label{fig:obst}
  \Description[Horizontal bar charts indicating notification type preferences: push, no feedback, and pre-setting. Each row begins with a variable relating to the obstacle category, followed by counts of respondents selecting notification types.]{
Obstacle Notification        Push    No Feedback    Pre-Set
Display                      7       0              3
Trash Can                    4       5              1
Planter                      3       7              0
Column                       3       7              0
Floor Change (Carpet/Tile)   2       8              0
Pedestrian                   2       3              5
Steps/Stairs/Escalator/Ramp  6       0              0
Dropped Item                 2       0              0
}
  
\end{subfigure}

\begin{subfigure}{1\linewidth}
  \centering
   \includegraphics[width=1\linewidth]{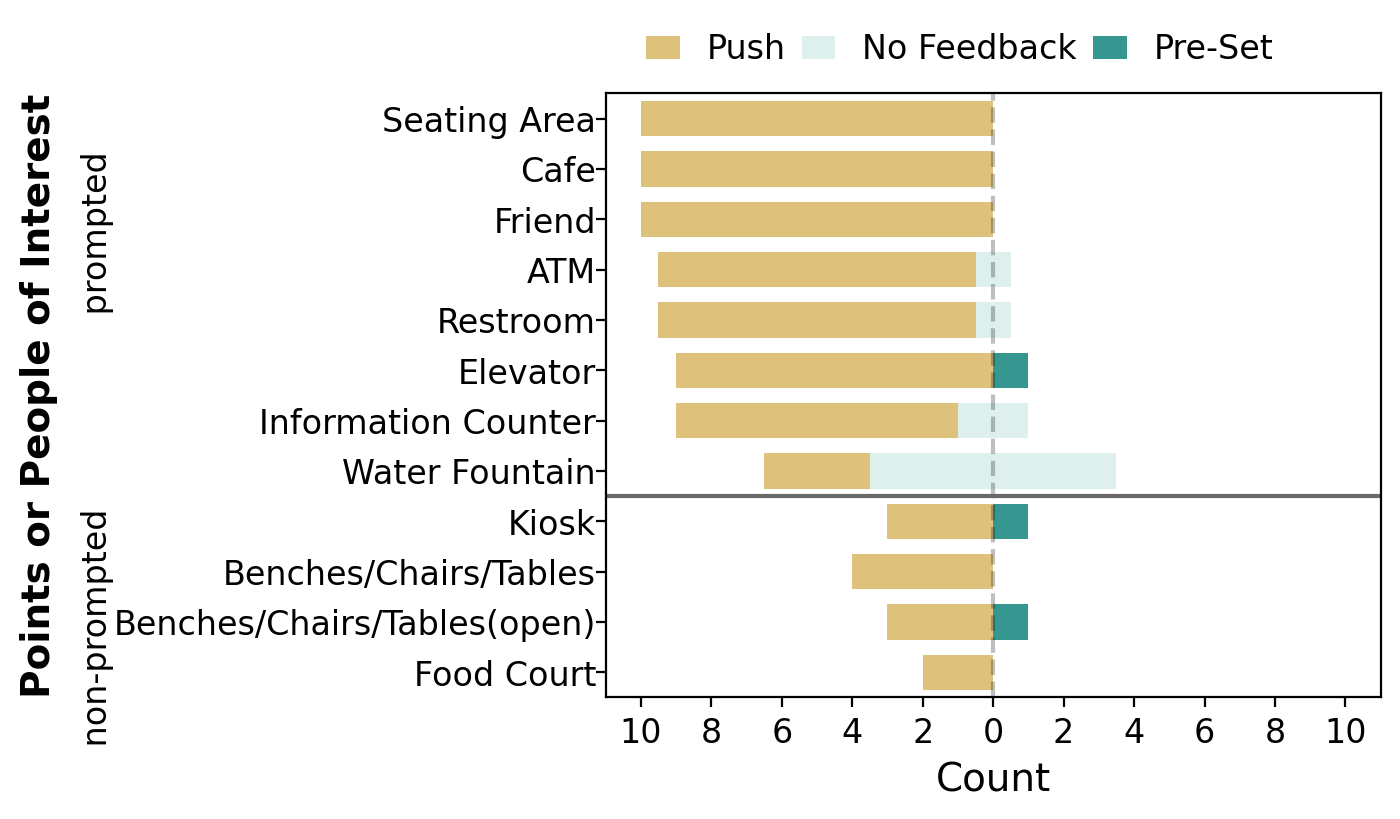}
  \caption{}
      \Description[Horizontal bar charts indicating notification type preferences: push, no feedback, and pre-setting. Each row begins with a variable relating to the points of interest category, followed by counts of respondents selecting notification types.]{
POI Notification             Push    No Feedback    Pre-Set
Seating Area                 10      0              0
Cafe                         10      0              0
Friend                       10      0              0
ATM                          9       1              0
Restroom                     9       1              0
Elevator                     9       0              1
Information Counter          8       2              0
Water Fountain               3       7              0
Kiosk                        3       0              1
Benches/Chairs/Tables        4       0              0
Benches/Chairs/Tables(open)  3       0              1
Food Court                   2       0              0
}
  \label{fig:poi}
\end{subfigure}
\vspace{-2em}
\caption{Types of desired notifications for O\&M including obstacles and POIs. Bars represent the frequency of responses from 10 blind participants, indicating their preferences regarding push/pre-set notifications or no feedback given a list of prompted features. We also include counts for non-prompted features that participants added to indicate their specific interests.}
\label{fig:OMNot}
\end{figure}

We expand on participants' responses to features of the surroundings that can serve as important information for orientation and mobility (O\&M), including obstacles and POIs. Figure~\ref{fig:OMNot} shows their preferences for rendering different types of O\&M notifications.

\begin{description}[style=unboxed,leftmargin=0.5cm]

\item[Non-functional obstacles \eg, \textnew{columns and planters}.] 
Aligned with the insights from our focus groups, many participants (except P9, P15, P16) preferred to receive no notifications about \textit{non-functional} obstacles such as columns in the environment if a guide robot is navigating around them. Yet again, these preferences tend to be impacted by mobility aids, as illustrated by P9: \textit{``I'm a cane user by default. I'm looking at my environment by cane.''} Some participants using a white cane wished for push (P9, P16) or pre-set notifications (P15) about columns serving as cues to better understand their current location and surroundings. Indeed, real-world trajectory data from blind users in prior work~\cite{kacorri2018environmental} revealed that information about surrounding obstacles improves path-following behavior. Similarly, while many participants (except P9, P16, P17) found the notifications about planters unnecessary, the rest of the white cane users preferred push notifications. 



\item[Functional obstacles \eg, \textnew{trash cans and displays}.] 
We found varying preferences for notifications considering how some obstacles could be \textit{functional} similar to our focus group insights. While half of the interview participants (P10, P11, P12, P14, P18) indicated no interest in knowing about trash cans in the walkways, the other half favored push notifications (P9, P15, P16, P17) for trash cans or pre-set notifications with toggle controls (P13). In addition, half of the participants (P10, P11, P13, P16, P18), including those who showed no initial interest in being notified, looked for a feature to search for trash cans in the mall and request directional guidance.

\vspace{0.5mm}




Participants also considered displays as functional obstacles, with many (except P11, P12, P13) preferring push notifications for information that could be shown on these displays---\eg, indoor maps or advertisements. The remaining three participants favored pre-set notifications instead for greater control over the information they would receive, especially when passing by multiple displays in the environment.

\item[Low-level obstacles \eg, \textnew{floor changes and steps}.]
Only a few participants (P13, P16) preferred push notifications about floor changes such as carpets or titles. Aligned with our focus group insights, the rest of the majority showed no interest in such information as they could notice these changes on their own. Instead, push notifications were deemed preferable for addressing low-level obstacles only when they posed ``significant trip hazards'' (P14) or were of a ``safety-oriented'' nature (P11). As examples of this nature, participants referred to steps (P10, P11, P14, 16), stairs (P11, P12, P15), and ramps (P10). A few (P9, P12) were also concerned about small items that people have dropped on the floor as unexpected objects, alerting the need for push notifications. 




\item[People \eg, \textnew{pedestrians and friends}.] 

When asked about potential obstacles, all participants referred to pedestrians. However, relevant notifications for pedestrians differed across participants. Half of the participants (P12, P13, P14, P15, P17) mentioned the importance of push notifications only in particular scenarios when heightened awareness would be crucial. This included navigating through crowded spaces or being near children or those using a wheelchair. 
\vspace{0.5mm}

Among the remaining half, two participants (P9, P16), both white cane users, preferred push notifications during navigation around pedestrians, whereas three participants (P10, P11, P18), including two guide dog owners, preferred no notifications. Much like other obstacle notifications, blind people's current navigation methods may shape their preferences. However, it is crucial to consider the potential long-term effects of using guide robots, particularly as an alternative to traditional mobility aids~\cite{guerreiro2019cabot}. Preferences for obstacle notifications could change over time and context, as highlighted by P9 a white cane user: \textit{``For people who use a guide dog, they are already used to where guide dogs will go around obstacles...a person or trash can. That is something I am not used to. That would take me time to do.''}

\vspace{0.5mm}

Pedestrian avoidance also highlighted instances involving specific individuals of interest. All participants agreed to receive push notifications when approached by friends, suggesting a guide robot's capability to detect pedestrians in such scenarios as illustrated by P15:  \textit{``It has happened to me on occasion where I'll be at a mall and someone recognizes me and comes up to me that knows me and stops me and says hello. The robot might think to turn as obstacles.''}



\item[POIs \eg, \textnew{kiosks, elevators, and seating areas}.]

As illustrated in Figure~\ref{fig:poi}, the relevance of POIs is aligned with what is found in navigation apps for blind people~\cite{sato2019navcog3,guerreiro2019airport}. All or most participants preferred push notifications for information not unique to window shopping scenarios, such as information counters, ATMs, restrooms, cafes, or elevators. Some participants suggested kiosks (P13, P14, P15, P17) and food court vendors (P10, P18) as additional POIs relevant to shopping mall environments.

\vspace{0.5mm}

All participants also expressed a preference for push notifications to know nearby seating areas. Yet, there were many mixed comments on the desirable methods for receiving relevant information in more dynamic scenarios---\eg, being notified only when benches, chairs, and/or tables were available (P9, P10, P13) or toggling notifications about their availability (P15). A few participants also suggested a search feature for available or closest benches (P11, P14) to avoid repeated notifications in the walkways. Interestingly, P14 thought of integrating pedestrian detection to gauge seat availability in the areas.

\end{description}


\subsection{Types of Shop Notifications}
\label{notifyshops}

\begin{figure}[t]
  \centering
  \includegraphics[width=0.85\linewidth]{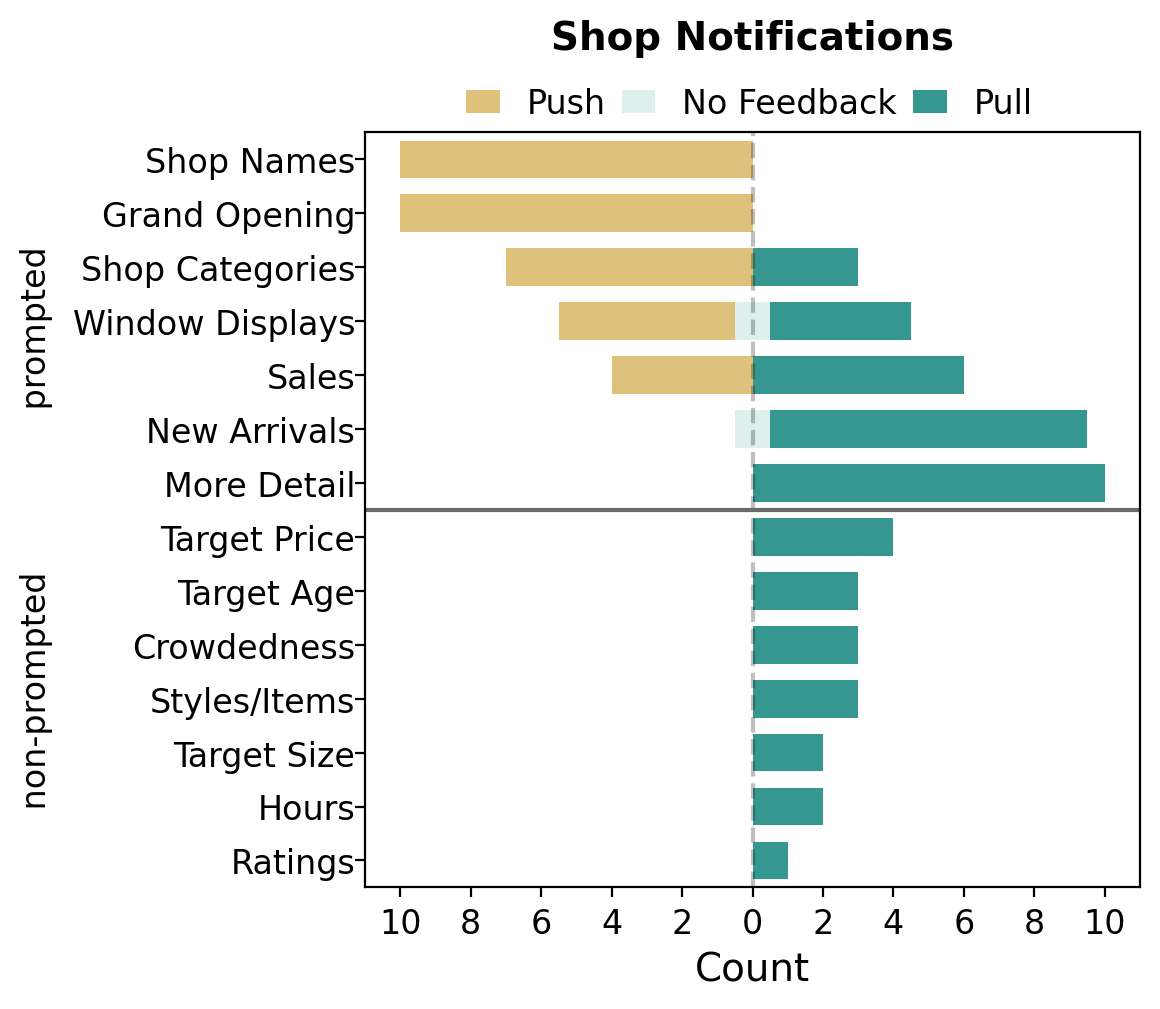}
  \caption{Types of desired notifications for shop information. Bars represent the frequency of responses from 10 blind participants, indicating their preferences regarding push/pull notifications or no feedback given a list of prompted shop-related features. Counts for non-prompted features indicate their specific interests.}

  \Description[Horizontal bar charts indicating notification type preferences: push, no feedback, and pull. Each row begins with a variable relating to the shop information, followed by counts of respondents selecting notification types.]{
Shop Notification    Push    No Feedback    Pull
Shop Names           10      0              0
Grand Opening        10      0              0
Shop Categories      7       0              3
Window Displays      5       1              4
Sales                4       0              6
New Arrivals         0       1              9
More Detail          0       0              10
Target Price         0       0              4
Target Age           0       0              3
Crowdedness          0       0              3
Styles/Items         0       0              3
Target Size          0       0              2
Hours                0       0              2
Ratings              0       0              1
}
~\label{fig:ShopNot}
\end{figure}

We investigate how blind participants' preferences for shop notifications vary across different layers of information in the context of window shopping (Figure~\ref{fig:ShopNot}). As information depth increases, it becomes crucial to design feedback solutions that do not overload users. We identify relevant information that participants want to access for more in-depth details and further probe their preferences to control the desired level of information.

\begin{description}[style=unboxed,leftmargin=0.5cm]

\item[First layer - \textnew{shop names, categories, and window displays}.] Aligned with the focus group discussions, all interview participants preferred to be informed of each shop they passed by, with a suggested format like \textit{``[Name of the shop] on your right''} included in a first layer of basic information. In addition, the majority of participants (except P10, P11, P15) wanted to know the categories (\eg, men's, women's, or teen's fashion, shoes) along with the names of shops via push notifications. The remaining few participants wanted to pull such short descriptions only for the shops of interest for better control.

\vspace{0.5mm}

Similar to what we observed in the focus group discussions, the interview participants showed a bimodal reaction in terms of information about window displays in nearby shops. Half of them (P9, P14, P15, P16, P18) expressed a desire for push notifications detailing these window displays in the first layer of information---an element that tends to be missing in their shopping experiences, as highlighted by P15: \textit{``I would love to hear that sort of thing [visual description]. That is the whole element that I miss when I go by myself. I like to hear those details, what the mannequins are wearing, like cream-colored top.''} Such preferences, when the remaining half did not want the push notifications, point to the need for customization or easy toggles between the first and second layers of information.





\item[Second layer - \textnew{more details}.] All participants agreed to pull more detailed information on an ``as-needed basis,'' often for the shops they are interested in to complement the information in the first layer. Their additional comments also surfaced diverse needs for relevant secondary information, including target age groups (P10, P15, P18), target size groups (P10, P18), possible price ranges (P11, P13, P14, P16), shopper ratings (P13), store hours (P9, P14), crowded or waiting lines (P12, P16, P18), general styles or items carried (P14, P15, P17), and window displays (P11, P12, P13, P17).


\vspace{0.5mm}

Some also shared their ideas on labeling shops for contextual information. P13 suggested categorizing shops as high-end, mid-range, low-end, or discount stores. P14 and P15 referred to the styles of fashion (\eg, dresses, casual/formal, business attire, outerwear, sleepwear, weekend wear) to complement commonly available shop tags of Men's or Women's fashion. 



\item[Promotional layer - \textnew{sales, new arrivals, and grand opening}.] When asked about preferences for notifications about promotional information extracted from our focus groups, interview responses varied based on the nature of the information and individual interests. While the majority preferred to pull more detail for selected shops for sales or new arrivals, some participants (P10, P12, P13, P18) preferred push notifications, especially for sales being considered in the first layer of information. P10 described how \textit{``having promotion comes in handy''} in determining whether to go inside stores and explore what's new, as \textit{``there could be something out there that you've never thought of.''} Perhaps for this reason, all participants agreed that they favor push notifications for grand opening signs. 







\end{description}

\subsection{Interaction Modalities}

To address the information requirements of blind people to enhance window shopping opportunities, we are interested in their preferences to interact with a future guidance system, including the design of pull, pre-set, and search features. We explore user reactions to various notification formats and interaction modalities. 

\begin{description}[style=unboxed,leftmargin=0.5cm]
\item[Pull features - \textnew{button, voice controls, and mobility patterns}.] 

When asked about how they would prefer to pull information about shops and other features of the surroundings, many participants (P9, P10, P11, P13, P14, P15) referred to active methods of using buttons (\eg, on a guide robot's handle) to navigate through shop details. For example, P15 imagined a scenario of having a short press of a 'more information' button for shop categories and a long press for detailed information about promotions. A few participants (P16, P18) also suggested voice commands to request specific information, such as `What are the sales?' The remaining participants (P12, P17) referred to passive methods; if they stop, slow down, or dwell after they hear the shop names, their movements can trigger to pull detailed shop information.  

\vspace{0.5mm}

Additionally, P11 articulated the scaffolding of information, particularly with ways to skip or cut off some layers of information: \textit{``It would be nice though if there was an option to skip through to specifically go to what I wanted to know, like if I want to know if the shop carries formal wear, that would maybe be like the third thing it says and I might want to skip the second thing. Having that as an option or being able to cut it off maybe like as it starts to say `Banana Republic' and `tell me more' and 'this is a clothing store' and I can just be skip skip.''}

\item[Pre-set features - \textnew{settings and toggle controls}.] Some participants (P12, P13, P18) expanded on how they want to control the types of information for pre-set notifications. They referred to customizing with toggle controls via a smartphone application, as illustrated by P13: \textit{``I would probably make it like a setting, like swipe through an iPhone and toggle things that I want. Pedestrians, I wouldn't necessarily need to know. It might be something that is a choice for the user.''} P12 and P18 also suggested a setting to specify 'my favorite shops' or 'my favorite types of shops' to tailor push notifications such as promotional information only for the shops under these predefined categories.




\item[Search features - \textnew{voice queries and physical interactions}.] Most participants (except P9) preferred to search for specific objects in the environment by voice. P9 felt more comfortable using a menu on a smartphone application. The majority still preferred voice commands to allow their hands to be free as they could be using a white cane or in a wheelchair. Some participants (P10, P15, P16, P17) also referred to how they might have shopping bags in their hands, leading to cumbersome tasks to take out and control with their smartphone. Yet, we could expect privacy concerns about voice-search features, as illustrated by P11: \textit{``I would want to have a voice activated feature just so because that seems like it's the easiest but I think also have an another option would be nice. I wouldn't be embarrassed to say `Take me to the restroom' but I think other people might be when they don't want to say that out loud in a crowd. Another option is a physical thing with a smartphone, something more discrete.''}

\end{description}


\section{Discussion}

Our research serves as a step toward supporting blind people in recreational shopping, particularly focused on window shopping (\eg, exploration and serendipitous discovery in shopping malls), which can be highly conducive to one's leisure and social participation~\cite{baker2006consumer,baker2002can}. First, through our focus groups with 8 blind participants, we explored the information needs, challenges, and current approaches they have to window shopping. Engaging in such activities was found to be limited, primarily due to a lack of independent access to information. Consequently, they often employed strategies to minimize exploration costs, relying on assistance from store staff and/or friends to effectively search for specific stores and items. 

Following our focus groups, we proceeded to conduct one-on-one interviews with 10 additional blind participants and delved deeper into the analysis of requirements to enhance recreational window shopping opportunities. Below, we reflect on the information requirements identified and translate our findings in the context of AI-infused systems and more recent work on generative AI.

\subsection{Design Implications for Information}

Both focus groups and requirements analysis interviews gave us insights into the interface design to navigate through different types of information considering the diverse needs. For example, the relevance of O\&M notifications was dependent on many personal and situational factors, including the usage of aids for mobility and the type of obstacles (\eg, functional, non-functional). In the case of white cane users, receiving notifications about obstacles, even for non-functional objects such as columns, can serve as \textbf{helpful cues for orientation}. Indeed, prior work on real-world data indicates that informing participants about surrounding obstacles improves path-following behavior~\cite{kacorri2018environmental}. Furthermore, functional obstacles such as trash cans or chairs can become \textbf{features of interest when needed}. Recognizing these factors, there is a clear need to facilitate the customization of notifications, allowing users to preset their preferences and toggle features of interest for an adaptive notification design. 

The relevance of notifications regarding pedestrians in the surroundings was found to be context-dependent, with only a few participants from the interviews considering this information a higher priority via push notifications. Our analysis further revealed situational characteristics to be integrated into pedestrian detection, such as the presence of \textbf{crowds, children, or wheelchair users}. Surprisingly, till now, the discourse on pedestrian detection for accessibility has focused on detecting individuals (not crowds), adults (not children), and demographics (not mobility aids), though we see recent efforts like the AVA Data Challenge~\cite{cvpr2024ava} helpful in this area.  
The design components could also involve personalizing the recognition system~\cite{lee2021accessing}, as identifying \textbf{people of interest} (\eg, friends) was deemed critical by our participants in extending the ability of pedestrian detection. Perhaps for this reason, despite the extensive literature focusing on pedestrian detection for blind people, there remains a large gap between the research trends in this field and users' actual needs~\cite{gamage2023blind}. We suggest considering these contextual elements as potential avenues to close this gap. 

In terms of information requirements related to shops, it is important to take into account different levels of information-seeking behaviors. Our analysis showed how most participants preferred to receive push notifications for shop names as they walked past shops, along with short descriptions of shop categories (\eg, Women's fashion) and to have controls to pull more details about the shops. Yet, with preferences for shop-related details varying among participants, scaffolding techniques for shop notifications would be critical. For example, adapting to user preferences could be facilitated by \textbf{interactive verbosity controls} allowing users to skip certain information (P11), or \textbf{customization settings} to prioritize specific information \eg, crowdedness under `\textit{my favorite shops}' (P12, P18).

We also observed that visual promotional content such as window displays could play an important role in fulfilling recreational and informational purposes in shopping, as exemplified by P18: \textit{``What the display is trying to promote, that could be a fashion statement that somebody is making. If it's for someone who sees it, it just bounces in their head. `Oh wow the beach, that swimsuit,' and they say `I think I'll go in and check that shop out.' How do you make that something alluring to me.''} A great number of techniques have leveraged computer vision to automate the process of producing textual descriptions of visual content~\cite{hanley2021computer,wu2017automatic,gurari2020captioning}, but the results can be limited, especially in \textbf{providing `\textit{alluring}' information}. One future direction could involve exploring approaches to generate contextual descriptions, such as demonstrated by Sreedhar \etal~\cite{sreedhar2022aide} to create context-specific alt-text for online retail images. In the setting of window shopping, incorporating review comments and input from store employees designing window displays or communicating the `fashion statement' might be valuable for meaningful descriptions. With advances in generative AI, there is potential to generate narrative and interpretive descriptions that can pique curiosity, similar to work being done in storytelling~\cite{uehara2022vinter,ravi2021aesop}.


\subsection{Design Implications for Interactions in O\&M}
Based on the findings, we propose the following user interface designs and considerations to enhance existing wayfinding and navigation systems to support blind recreational window shopping.

\textbf{Navigation apps in smartphones.} Turn-by-turn navigation apps for blind people have great potential to be used in window shopping scenarios. In fact, the evaluation study of NavCog~\cite{sato2017navcog3} has surfaced a design suggestion for a ``window shopping mode'' that enables more frequent notifications related to nearby shops. Going forward, it is important to address the means to navigate the hierarchy of shop information, as our findings highlight the first (\eg, shop names and categories) and second (\eg, sales and new arrivals) layers. One way is to incorporate gesture-based interaction, which has been discussed by Kacorri \etal~\cite{kacorri2018insights} in the context of a mobile interface for blind users to request information while in motion. We can extend this direction based on our participants' feedback, such as a long press on the touchscreen to activate the second layer. This method can also be complemented with existing swiping gestures (\eg, left and right swipe to interact with POIs on a virtual route for blind people~\cite{guerreiro2017virtual}) to access a series of detailed shop information.

One opportunity we anticipate lies in the ability to incorporate data about shops. When designing the Wizard-of-Oz app as outlined in the original protocol (refer to Section~\ref{reflections}), we did this step \textit{manually}. We referenced floor plans and a comprehensive list of shops on the shopping mall's website which included annotations for shop categories and merchandise types. The official shop website and social media were also used to collect promotional information. Clearly, such a manual approach is not scalable and struggles to deliver relevant, up-to-date information. Yet, the labor-intensive process of collecting and annotating POIs is not a surprise in the implementation of navigation apps~\cite{guerreiro2019airport,sato2017navcog3}. We see an opportunity with the recent advancements in generative AI \eg, by combining large language models with real-time search~\cite{davis2023chatgpt} or augmenting them with internal knowledge bases~\cite{ram2023context} specific to particular shopping malls.


\textbf{Guide robots.} Building on existing robot guidance systems, such as Cabot~\cite{guerreiro2019cabot}, we suggest future scenarios to enhance window shopping opportunities. Blind users can physically hold the robot to navigate around obstacles, potentially offering benefits as a `robotic guide dog'~\cite{guerreiro2019cabot,hwang2024towards}. Importantly, by incorporating various sensors, cameras, and machine learning algorithms to detect features of the surroundings, guide robots can be designed to provide relevant information for recreational shopping and communicate them with the user not just with through audio but with haptic feedback. 

As informed by our study, the handle can be embedded with physical buttons for primary and secondary actions to access different layers of shop information. Additionally, it can be equipped with a motion command interface, requesting notifications as they slow down or stop to explore the nearby shops of interest. Our participants also suggested integrating the robot with a smartphone app to enable customization, such as pre-setting their favorite shops, to change the priority of information (\eg, push notifications for sales only for the pre-defined shops). Furthermore, a voice command for searching POIs, such as seating areas or trash cans, could be an option for the user, as most participants preferred to limit the cumbersome tasks associated with taking out and using their smartphones in combination with guide robots. Yet, as reflected on P11's comments \textit{``I wouldn't be embarrassed to say `Take me to the restroom' but I think other people might be when they don't want to say that out loud in a crowd,''} carefully considering social and cultural contexts is critical.

Social acceptability related to technology use in recreational window shopping warrants further design considerations. Our focus groups revealed societal barriers to recreational shopping opportunities for blind people, encapsulated by P6 referring to \textit{``sighted people interference.''} This reflects blind people's concerns discussed in previous research~\cite{azenkot2016enabling}, emphasizing the need for design decisions that can mitigate possible social tensions arising from the interaction with guide robots. Perhaps we could learn from the recent effort by Cai \etal~\cite{cai2024navigating} considering social conventions in the design of a robotic guidance system, such as matching its speed with nearby pedestrians or maintaining an appropriate distance from walls.

\section{Conclusion}

Through a combination of focus groups and requirements analysis interviews with 18 blind participants, we identified preferences and requirements for information access to support blind people's recreational window shopping opportunities. We discerned relevant push notifications, which included visually engaging content like window displays, and user interfaces to manage the delivery of information related to obstacles and POIs. Customization of notifications was deemed necessary due to personal and situational factors, such as one's mobility aid used or the presence of crowds.  

In this paper, we discussed the implications and potentials for generative AI in the context of eliciting information as well as incorporating recreational window shopping in guide robots and navigation apps for the blind. More so, we discussed the broader social context surrounding these opportunities, as participants articulated their negotiation tactics to limit the social costs of in-person shopping, such as searching for information beforehand and seeking assistance from trained store staff. These discussions resonate with ongoing discourse on the future of shopping, connecting online platforms to an elevated in-person experience~\cite{moore2023future,weintraub2020future}---\eg, advanced support from sales associates and technology to enrich human interaction. 
Thus, to create more inclusive in-person recreational experiences for blind people, we aim next to delve into social navigation aspects~\cite{lee2021accessing, francis2023principles} within guidance systems.

\begin{acks}
We would like to thank the blind participants in our study as well as Daisuke Sato for helping us prepare the CaBot demo. This work was sponsored by Shimizu Corporation. Rie Kamikubo and Hernisa Kacorri were additionally supported from the National Institute on Disability, Independent Living, and Rehabilitation Research (NIDILRR), ACL, HHS (grants \#90REGE0008 and \#90REGE0024).
\end{acks}

\bibliographystyle{ACM-Reference-Format}
\bibliography{manuscript}

\appendix

\end{document}